\documentclass[useAMS]{mn2e}

\usepackage{times}
\usepackage{graphicx}
\usepackage{amsmath}
\usepackage{txfonts}
\newcommand{\beq}{\begin{equation}}
\newcommand{\bea}{\begin{eqnarray}}
\newcommand{\eeq}{\end{equation}}
\newcommand{\eea}{\end{eqnarray}}

\title[MHD-instabilities at ultra-relativistic magnetized shock
  waves]{On Fermi acceleration and MHD-instabilities at
  ultra-relativistic magnetized shock waves}

\author[G. Pelletier, M. Lemoine and A. Marcowith]
{Guy Pelletier$^{1}$\thanks{e-mail:{\tt
      guy.pelletier@obs.ujf-grenoble.fr}},
Martin Lemoine$^{2}$\thanks{e-mail:{\tt lemoine@iap.fr}} and
Alexandre Marcowith$^{3}$\thanks{e-mail:{\tt
    alexandre.marcowith@lpta.in2p3.fr}}\\
$^{1}$ Laboratoire d'Astrophysique de Grenoble, \\
	CNRS, Universit\'e Joseph Fourier II,
	BP 53, F-38041 Grenoble, France; \\
$^{2}$ Institut d'Astrophysique de Paris, \\ 
	UMR 7095 CNRS, Universit\'e Pierre \& Marie Curie, 
	98 bis boulevard Arago, F-75014 Paris, France\\
$^{3}$ Laboratoire de Physique Th\'eorique et Astroparticules,\\
      CNRS, Universit\'e Montpellier-II,
      place Eug\`ene Bataillon, 34095 Montpellier C\'edex, France}

\begin{document}

\date{}

\pubyear{2008}

\maketitle

\label{firstpage}

\begin{abstract}
Fermi acceleration can take place at ultra-relativistic shock waves if
the upstream or downstream magnetic field has been remodeled so that
most of the magnetic power lies on short spatial scales. The relevant
conditions under which Fermi acceleration become efficient in the
presence of both a coherent and a short scale turbulent magnetic field
are addressed. Within the MHD approximation, this paper then studies
the amplification of a pre-existing magnetic field through the
streaming of cosmic rays upstream of a relativistic shock wave. The
magnetic field is assumed to be perpendicular in the shock front
frame, as generally expected in the limit of large shock Lorentz
factor. In the MHD regime, compressive instabilities seeded by the net
cosmic-ray charge in the shock precursor (as seen in the shock front
frame) develop on the shortest spatial scales but saturate at a
moderate level $\delta B/B \sim 1$, which is not sufficient for Fermi
acceleration. As we argue, it is possible that other instabilities
outside the MHD range provide enough amplification to allow successful
Fermi acceleration.
\end{abstract}


\section{Introduction}
The physics of collisionless shock waves is of paramount importance in
modern astrophysics, as it governs a variety of phenomena, most
notably the emission of high energy radiation. In particular, the
afterglow radiation of gamma-ray bursts is generally attributed to the
synchrotron emission of electrons that have been accelerated around
the relativistic external shock wave, with shock Lorentz factor
$\Gamma_{\rm sh}\,\sim\,100$ (Paczy\'nski \& Rhoads 1993, Katz 1994,
M\'esz\'aros \& Rees 1997, Sari \& Piran 1997, Vietri 1997, Waxman
1997; see Piran 2005 for a review). In this framework, it can be shown
that the magnetic field, downstream of the shock wave, must have been
amplified by orders of magnitude beyond the shock compression of a
typical interstellar magnetic field (see for instance Waxman 1997; see
also Piran 2005 and references therein). Recent work by Li \& Waxman
(2006) further shows that even the upstream magnetic field must have
been amplified by at least one or two orders of magnitude in order to
explain observed $X-$ray afterglows.

These results should be put in perspective with recent studies on the
process of Fermi acceleration in the test particle limit around
relativistic shock waves. In particular, Niemiec \& Ostrowski (2006)
have shown that Fermi acceleration becomes inoperative in a turbulent
magnetic field with power spectra of the Kolmogorov or scale invariant
type. Through analytical calculations, Lemoine, Pelletier \& Revenu
(2006) have shown that relativistic Fermi acceleration is inefficient
if the magnetic field power is distributed on scales larger than the
typical Larmor radius of the accelerated population, in agreement with
the above numerical result. Therefore the interpretation of the
gamma-ray burst afterglow emission from the relativistic external
shock wave requires the magnetic field to have been amplified on short
spatial scales.

It has been suggested that the relativistic two stream Weibel
instability could amplify the downstream magnetic field to the level
required by afterglow modeling of gamma-ray bursts (Gruzinov \& Waxman
1999; Medvedev \& Loeb 1999). Subsequent studies have however argued
that this instability should saturate early on (Wiersma \& Achterberg
2004; Lyubarsky \& Eichler 2006). The question of the long term
evolution of the downstream magnetic field (on timescales
$\gg\,\omega_{p}^{-1}$, with $\omega_{p}$ the plasma frequency) also
remains open. Ongoing particle-in-cell simulations should eventually
shed light on this issue (Silva {\it et al.} 2003; Frederiksen {\it et
  al.} 2004; Medvedev {\it et al.} 2005; Kato 2007; Chang, Spitkovsky
\& Arons 2008; Keshet {\it et al.} 2008; Spitkovsky 2008).

Regarding the amplification of the upstream magnetic field, the
present situation is reminiscent of results obtained for
non-relativistic supernovae remnant shock waves, where the
interstellar magnetic field has apparently been amplified by one or
two orders of magnitude (see V\"olk, Berezhko \& Ksenofontov 2005,
Parizot {\it et al.} 2006).  In this case, the leading candidate for
the instability is the streaming instability, seeded by the cosmic ray
precursor in the upstream plasma (see Bell 2004, 2005; Pelletier,
Lemoine \& Marcowith 2006, Marcowith, Lemoine \& Pelletier 2006;
Reville {\it et al.} 2007; Niemiec et al. 2008; Amato \&
Blasi 2008; Zirakashvili, Ptuskin \& V\"olk 2008; Reville et al. 2008). The generalization
of this instability to the relativistic regime has been studied on
phenomenological grounds by Milosavljevi\'c \& Nakar (2006) who have
concluded that it should be able to account for the degree of
amplification inferred in gamma-ray bursts. More recently, Reville,
Kirk \& Duffy (2006) have derived in detail the dispersion relation
for a parallel shock wave and the saturation due to thermal effects.

In the present work, we propose to explore in detail the
generalization of this type of instability to the ultra-relativistic
regime $\Gamma_{\rm sh}\,\gg\,1$. One crucial difference with the
previous works on the streaming instability is that we consider the
most natural case of superluminal shock waves (with a magnetic field
perpendicular to the shock normal in the shock front frame). This case
is more generic than the parallel configuration studied previously
because the transverse component of the magnetic field is boosted by
the shock Lorentz factor when going to the shock frame. Another
important difference is that we bring to light a new type of
instability, of a compressive nature. Finally, in contrast with most
particle-in-cell simulations performed to date, our study focusses on
magnetized shock waves, for which there exists a coherent upstream
magnetic field (whose dynamical influence on the shock jump conditions
can be neglected however). Nevertheless there exist pioneering PIC simulations in the moderately relativistic regime, which include both a mean field and a significant mass ratio between electrons and ions, see Hededal \& Nishikawa (2005), Dieckmann, Shukla, \& Drury (2008).

We adopt a simplified description in which the cosmic-ray distribution
is modeled as a step function out to some distance $\ell_{\rm cr}$ and
we neglect the cosmic-ray response to the disturbance. This latter
assumption is justified by the fact that the instability is maximal on
the shortest spatial scales, orders of magnitude below the typical
Larmor radius of accelerated particles.
  
The paper is organized as follows. In Section~\ref{sec:general}, we
introduce the main scales of the problem, most notably the diffusion
scale of the cosmic rays; we then calculate the level of amplification
that is necessary to make Fermi acceleration
operative. Section~\ref{sec:instab} is devoted to the investigation of
the instabilities under the condition that some cosmic-rays have
undergone a first Fermi cycle. Section~\ref{sec:conc} summarizes out
results and provides some outlook.  Details of the calculations are
provided in Appendix~\ref{sec:appperp}.

\section{General considerations}\label{sec:general}
We carry out most of the discussion in the shock front rest frame,
hence unless otherwise noted, all quantities are evaluated in this
frame. We use the subscripts $_{\rm\vert u}$ or $_{\rm\vert d}$ to tag
quantities measured in the upstream or in the downstream rest frame
respectively.

\subsection{Upstream diffusion length}
\label{sec:updiff}
In the upstream rest frame, cosmic rays can never stream too far ahead
of a relativistic shock wave since this latter propagates towards
upstream with velocity $v_{\rm sh}=\beta_{\rm sh}c\approx c$ [the
  shock Lorentz factor $\Gamma_{\rm sh} \equiv (1-\beta_{\rm
    sh}^2)^{-1/2}\,\gg\,1$].  Cosmic rays scatter on magnetic
turbulence upstream before they are caught back by the shock wave when
their pitch angle $\theta_{\rm\vert u} \sim 1/\Gamma_{\rm sh}$
(Gallant \& Achterberg 1999, Achterberg {\it et al.}
2001). Consequently, they can travel a distance $\ell_{\rm cr\vert
  u}$, which may take the following values depending on the ratio of
the Larmor radius $r_{\rm L\vert u}$ to the coherence length of the
upstream magnetic field $\lambda_{\rm c\vert u}$ (Milosavljevi\'c \&
Nakar 2006):
\begin{itemize}
\item for small scale turbulence 
\begin{equation}
\ell_{\rm cr\vert u}\,\sim\,\frac{1}{\Gamma_{\rm sh}^2}\frac{r_{\rm
    L\vert u}^2}{\lambda_{\rm c\vert u}} \quad \left(r_{\rm L\vert
  u}\gg \Gamma_{\rm sh} \lambda_{\rm c\vert u}\right) \ ,
\end{equation}
\item for large scale turbulence
\begin{equation}
\ell_{\rm cr\vert u}\,\sim\,\frac{r_{\rm L\vert u}}{\Gamma_{\rm sh}} \quad \left(r_{\rm L\vert
  u}\ll \Gamma_{\rm sh} \lambda_{\rm c\vert u}\right) \ .
\end{equation}
\end{itemize}

Both regimes, short or large scale turbulence can be expected at some
point, insofar as the excitation of the upstream magnetic field on
short spatial scales is due to the streaming of the non-thermal
particle population in the shock precursor. Indeed, cosmic rays of the
first generation are to interact with a turbulent magnetic field
ordered on large scales. However, provided the instability that they
trigger grows fast enough, cosmic rays of the next generation will
propagate in short scale turbulence. In reality, the situation is
likely to be more complex as the process of particle propagation
upstream and magnetic field generation are closely intertwined. The
fact that the non-thermal population contains particles of different
energies, which can stream at different distances from the shock
front, should also play a significant role. In this respect, Keshet
{\it et al.} (2008) have observed that the upstream magnetic field is
affected to greater distances as time goes on. This strongly suggests
that higher energy cosmic rays are produced as time goes on, and that,
by travelling farther in the upstream medium, they excite the
turbulence at larger distances from the shock front.

In the discussion that follows, we estimate the growth of unstable
modes in both the limits of small or large scale turbulence in order
to remain as general as possible. Certainly the limit of large scale
turbulence is more restrictive with respect to the growth of the
instability, since the distance travelled upstream is significantly
reduced with respect to that in small scale turbulence. Whichever
limit prevails depends on the ratio of the short scale turbulent to
the large scale (coherent) magnetic field strength, as discussed in
Section~\ref{sec:fermi}. We also discuss the effect of higher energy
cosmic rays on the growth rate.

It is important to emphasize that the distance that controls the
growth of the instability is that between the shock front and the
position of the particle, which is smaller than $\ell_{\rm cr\vert u}$
by a factor $(1-\beta_{\rm sh})\sim \left(2\Gamma_{\rm
  sh}^2\right)^{-1}$. In the following, we will need the expression
for $\ell_{\rm cr\vert sh}$ (also noted $\ell_{\rm cr}$), i.e. the
length scale of the cosmic-ray distribution as measured in the shock
front rest frame. It can be calculated by transforming the upstream
residence time $t_{\rm r\vert u}\,\simeq\,\ell_{\rm cr\vert u}/c$ in
the shock front frame $t_{\rm r\vert sh}=t_{\rm r\vert u}/\Gamma_{\rm
  sh}$, and then by rewriting in the expression obtained the upstream
Larmor radius and coherence length in terms of their shock frame
equivalent. In the perpendicular (or superluminal and
ultra-relativistic) configuration of interest, $r_{\rm L\vert
  sh}\,\simeq\, \Gamma_{\rm sh}^{-2} r_{\rm L\vert u}$, $\lambda_{\rm
  c\vert sh}\,\simeq\, \Gamma_{\rm sh}^{-1} \lambda_{\rm c\vert
  u}$. This boost of the coherence length is valid for wavenumber
modes that are parallel to the shock normal; perpendicular wavenumber
modes remain unchanged.  In the following, we thus use the following
expressions for $\ell_{\rm cr}$:
\begin{itemize}
\item for small scale turbulence 
\begin{equation}
\ell_{\rm cr}\,\equiv\, \ell_{\rm cr\vert sh}\,\simeq\, \frac{r_{\rm L\vert
    sh}^2}{\lambda_{\rm c\vert sh}}\ ,
\label{eq:lbar1}
\end{equation}
\item for large scale turbulence
\begin{equation}
\ell_{\rm cr}\,\equiv\, \ell_{\rm cr\vert sh}\,\simeq\, r_{\rm L\vert
    sh}\ .
\label{eq:lbar2}
\end{equation}
\end{itemize}

Below, we find a process of generation of intense magnetic
disturbances at short spatial scales. The expected instabilities in
all relevant cases generate disturbances with a coherence length
$\lambda_{\rm c\vert u}$ between the minimal length $l_{\rm
  MHD}=c/\omega_{\rm p,i}$ required for the validity of MHD
description and the diffusion length $\ell_{\rm cr\vert u}$ of the
cosmic rays, but with a preference for short scale, i.e. $l_{\rm
  MHD}\,\la\,\lambda_{\rm c\vert u}\,\la\,\ell_{\rm cr\vert u}$.  The
minimal MHD length is:
 \begin{equation}
l_{\rm MHD} \equiv \beta_{\rm A} r_{0\vert\rm u} \ ,
\end{equation}
where $\beta_{\rm A} \equiv B_{0\vert\rm u}/\sqrt{4\pi \rho_{\rm
    u\vert u}c^2}$ denotes throughout this paper the Alfv\'en velocity
(measured upstream) in units of $c$, and $r_{0\vert\rm u}  =
m_{\rm p}c^2/eB_{0\vert\rm u}$ is the Larmor radius of thermal protons in
the upstream comoving frame. The above MHD scale is measured in the
upstream plasma rest frame.
 
\subsection{Requirements for successful Fermi acceleration}\label{sec:fermi}
As discussed in Lemoine, Pelletier \& Revenu (2006), a particle can
execute a large number of Fermi cycles through the shock only if the
turbulence coherence scale is much smaller than the Larmor radius
(assuming no coherent magnetic field), in detail $r_{\rm L\vert u}\gg
\lambda_{\rm c\vert u} \Gamma_{\rm sh}$ upstream, or $r_{\rm L\vert d}
\gg \lambda_{\rm c\vert d}$ as measured downstream. 

The Fermi process becomes inoperative in the limit of large coherence
length (when compared to the typical Larmor radius) since the cycle
time becomes smaller than a coherence time hence the field is
effectively regular and superluminal on a typical Fermi
cycle. Conversely, the Fermi process will be operative provided either
of the above inequalities is fulfilled, since the memory of the
initial conditions at shock crossing is then erased by pitch angle
diffusion in the short scale turbulence.

In the generic case of a ultra-relativistic superluminal shock, the
above two conditions are actually related by the shock jump
conditions, since the Larmor radius increases by $\sim \Gamma_{\rm
  rel}^2$ when going from downstream to upstream, while the coherence
length increases by $\sim\Gamma_{\rm rel}$. The quantity $\Gamma_{\rm
  rel}$ corresponds to the Lorentz factor of the downstream fluid as
measured upstream, and in the case of a ultra-relativistic strong
shock, $\Gamma_{\rm rel}\,\simeq\,\Gamma_{\rm sh}/\sqrt{2}$ (Blandford
\& McKee 1976).

In the following, we describe the upstream and downstream fluids as
comprising a large scale component $\mathbf{B_0}$ and a short scale
turbulent generated by some unspecified instability. In this case,
Fermi acceleration will be efficient provided the short scale field
has a sufficient amplitude as compared to $B_0$. We discuss this
condition in the following.

\subsubsection{Upstream motion}

\begin{figure}
  \centering{
    \includegraphics[width=0.4\textwidth,clip=true]{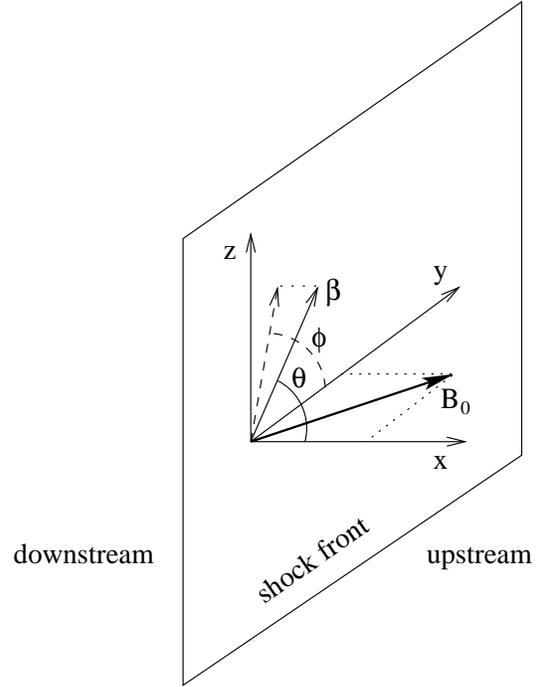}}
  \caption{Schematic plot of the geometrical configuration in the
    upstream plasma rest frame. The coherent component $\mathbf{B_0}$
    lies in the $x-y$ plane, with $x$ pointing towards the shock
    normal, and ($\theta$, $\phi$) the angles of the particle velocity
    in this frame.}
  \label{fig:shock_conf}
\end{figure}

In this particular section, all quantities are evaluated in the
upstream rest frame and we avoid using the notation $_{\rm\vert u}$
out of clarity. We derive here the requirements on the amplitude of
the short scale component of the magnetic field that would allow
successful Fermi acceleration. Without loss of generality, we assume
that the coherent component $\mathbf{B_0}$ lies in the plane $x-y$,
the $x$ direction corresponding to the shock normal (oriented toward
upstream infinity). We denote by $\mathbf{b}$ the irregular component
(whose amplitude is expressed in units of $B_0$). The layout is
pictured in Fig.~\ref{fig:shock_conf}.

The total field is thus written as $\mathbf{B}= B_0(\mathbf{e} +
\mathbf{b})$, and the particle trajectory is governed by the following
equation:
\begin{equation}
\frac{{\rm d}\boldsymbol\beta}{{\rm d}t} = \omega_{\rm L,0}\,
\boldsymbol\beta\times(\mathbf{e} + \mathbf{b})\ ,
\end{equation}
with $\omega_{\rm L,0}$ denoting the Larmor frequency expressed
relatively to $B_0$. We neglect the influence of any short scale time
varying electric field $\boldsymbol\delta\mathbf{E}$ in the above
equation of motion since $\delta E/(B_0b)\,\sim\, \omega/k\,\ll\,1$.

In order to quantify the return timescale and the condition for
successful Fermi acceleration as a function of the amplitude of the
random component of the magnetic field, it is useful to express this
equation in terms of the angles $\theta$ and $\phi$, which are defined
as: $\beta_x=\cos\theta$, $\beta_y=\sin\theta\cos\phi$ and
$\beta_z=\sin\theta\sin\phi$. These equations of motion read:
\begin{eqnarray}
\dot\theta&\,=\,& \omega_{\rm L,0}\,\left[\left(e_y +
  b_y\right)\sin\phi -
  b_z\cos\phi\right]\ ,\label{eq:t}\\ \dot\phi&\,=\,& \omega_{\rm
  L,0}\,\left[-e_x-b_x +
  \left(e_y+b_y\right)\frac{\cos\phi}{\tan\theta} +
  b_z\frac{\sin\phi}{\tan\theta}\right]\ .
\label{eq:p1}
\end{eqnarray}
Rescaling the Larmor frequency with respect to $e_yB_0$, i.e. defining
$\omega_{{\rm L},y}\,\equiv\, e_y\omega_{\rm L,0}$, and noting that
$\theta\,\ll\,1$ when the particle propagates upstream, the equations
of motion for $\theta$ and $\phi$ can be rewritten as follows:
\begin{eqnarray}
\dot\theta&\,\simeq\,& \omega_{{\rm L},y}\sin(\phi) + \omega_{{\rm
  L},y}A\sin(\phi-\sigma)\ ,\nonumber\\
\dot\phi&\,\simeq\,& \omega_{{\rm L},y}{\cos\phi\over\theta}\,+\,\omega_{{\rm
  L},y}A{\cos(\theta-\sigma)\over\theta}\ .
\label{eq:p2}
\end{eqnarray}
The factors $A$ and $\sigma$ describe respectively the amplitude and
the phase of the small scale magnetic field in the shock front plane,
i.e. $b_y=A\cos\sigma$ and $b_z=A\sin\sigma$ (assuming isotropic
turbulence). The amplitude $A=\delta B/B_y$ is measured with respect
to $B_y$.

One can extract from the above system the unperturbed trajectory,
i.e. assuming $A=0$:
\begin{equation}
\cos\phi\,=\,\cos\phi_{\rm i}\,{\sin\theta_{\rm i}\over\sin\theta}\ ,
\end{equation}
where values indexed with $_{\rm i}$ are calculated at some initial
time. Since entry into upstream corresponds to $\cos\theta_{\rm
  i}\,>\,\beta_{\rm sh}$ hence $\theta_{\rm i}\,\lesssim\,
1/\Gamma_{\rm sh}$, and exit from upstream corresponds to
$\theta\,\gtrsim\,1/\Gamma_{\rm sh}$, the above equation suggests that
$\phi$ is driven to an angle close to $\pm\pi/2$.  The sign is given
by the sign of $e_y$: for $e_y>0$, it is $+\pi/2$. This region lies
opposite to that which allows return to the shock when the particle
travels downstream (see Lemoine, Pelletier \& Revenu 2006), hence
Fermi cycles cannot be completed. It is important to note that this
unperturbed trajectory is executed on a timescale $t_{\rm
  unpert}\,\simeq\, r_{{\rm L},y}/(\Gamma_{\rm sh}c)$ (where $r_{{\rm
    L},y}=c/\omega_{{\rm L},y}$).

The noise term comes from the random phase $\sigma$ (assuming
isotropic turbulence in the shock front plane, an assumption that can
be relaxed). Therefore the trajectory deviates from the unperturbed
trajectory to linear order by $(\Delta\phi,\Delta\theta)$, whose
variance increases linearly with time:
\begin{eqnarray}
\langle\Delta\phi^2\rangle &\,\simeq\,& {1\over 3}A^2{\omega_{{\rm L},y}^2
  \over \theta^2}\tau_{\rm c}\Delta t\ ,\nonumber\\
\langle\Delta\theta^2\rangle&\,\simeq\,& {2\over 3}\,A^2\omega_{{\rm
  L},y}\tau_{\rm c}\Delta t\ .
\end{eqnarray}
Note that the above equation does not contain all the terms driving
the variance of $\phi$ and $\theta$: it neglects term of the form
$\int {\rm d}t_1{\rm d}t_2
\langle\Delta\phi(t_1)\Delta\phi(t_2)\rangle$ and similarly in
$\Delta\theta$. However, these terms are smaller by ${\cal O}(A^2)$ as
compared to those above, hence they can be neglected in a first
approximation (we will find $A\,\gg\,1$ further below).

In the absence of the regular driving term in the equations of motion
(i.e. when $B_y=0$), return occurs over a timescale $t_{\rm
  diff}\,\simeq\, \left(\Gamma_{\rm sh}^2A^2\omega_{{\rm L},y}^2\tau_{\rm
  c}\right)^{-1}$. Noise dominates over the unperturbed trajectory if
$\langle \Delta\theta^2\rangle\,\gtrsim\,\theta^2$, where $\theta$
indicates the unperturbed trajectory. This inequality must be
satisfied within a (return) timescale $t_{\rm unpert}$ otherwise the
particle will have exited from upstream (along the weakly perturbed
unperturbed trajectory) before the noise could overcome the
unperturbed driving force. This implies:
\begin{equation}
A^2\,\gtrsim\, {1\over \Gamma_{\rm sh}\omega_{{\rm L},y}\tau_{\rm c}}\ .
\end{equation}
If one defines the Larmor radius $r_{\rm L}$ measured with respect to
the total magnetic field, $r_{\rm
  L}\,=\,c\left[(A^2+1)^{1/2}\omega_{{\rm L},0}\right]^{-1}$, hence
$r_{\rm L}\,\simeq\, c/(A\omega_{\rm L,y})$ when $A\,\gg\,1$, the former
inequality amounts to:
\begin{equation}
A\,\gtrsim\, {r_{\rm L}\over \Gamma_{\rm sh}\lambda_{\rm c}}\ .
\end{equation}
If this inequality is satisfied, then one can check that
$\langle\Delta\phi^2\rangle^{1/2}\,\sim\,{\cal O}(1)$ on a return
timescale $t_{\rm diff}$, which implies that the return directions
are isotropized in the shock front plane. Fermi acceleration should
then be efficient (provided the return probability to the shock as
calculated downstream is also isotropic in $\phi$, see below).

An interesting implication of the above is to restrict Fermi
acceleration to a rather limited range of Larmor radii:
\begin{equation}
\Gamma_{\rm sh}\lambda_{\rm c}\,\lesssim\, \bar r_{\rm L}\,\lesssim\,
A\Gamma_{\rm sh}\lambda_{\rm c} \ .
\end{equation}
Very high values of the amplification factor are thus required to
produce powerlaw spectra over a large dynamic range. 

\subsubsection{Downstream motion}

All quantities in this subsection are evaluated in the downstream rest
frame. In this frame, the coherent magnetic field is nearly fully
aligned with the $y-$axis, and we assume indeed that
$\mathbf{B_0}=B_0\mathbf{y}$. It is useful to study the evolution of
the velocity components perpendicular ($\beta_\perp$) and parallel
($\beta_\parallel$) to the direction of $\mathbf{B_0}$:
$\beta_\parallel\,\equiv\,\beta_y$,
$\beta_\perp^2\,=\,\beta_x^2+\beta_z^2$.  The equations of motion for
the velocity $\mathbf{\beta}$ can then be written:
\begin{eqnarray}
\dot\beta_\perp&\,=\,&\omega_{{\rm
    L},0}{\beta_\parallel\over\beta_\perp}\left(\beta_xb_z -
\beta_zb_x\right)\,\nonumber\\ \dot\beta_\parallel&\,=\,&\omega_{{\rm
    L},0}\left(\beta_zb_x - \beta_xb_z\right)\ .
\end{eqnarray}
One can check that
$\beta_\parallel\dot\beta_\parallel+\beta_\perp\dot\beta_\perp=0$ as
it should since $\beta$ is conserved.  The unperturbed trajectory for
the couple of variables ($\beta_\parallel$, $\beta_\perp$) is trivial,
i.e. $\beta_{\perp,0}\,=\,{\rm const.}\,=\beta_{\perp,\rm i}$ and
$\beta_{\parallel,0}\,=\,{\rm const.}\,=\beta_{\parallel,\rm
  i}$. Return may occur or not, depending on these initial velocity
components; if it occurs, it does so on a timescale $t_{\rm
  unpert}\,\simeq\,r_{{\rm L},0}/c$ (Lemoine, Pelletier \& Revenu
2006).

Since $b_x$ is not compressed through shock crossing,
$b_x/b_z\,\sim\,1/\Gamma_{\rm sh}$ for isotropic upstream turbulence,
hence this component can be neglected. Integrating the above equations
to calculate the squared perpendicular displacement, one finds:
\begin{eqnarray}
\langle\Delta x_\perp^2\rangle&\,=\,&\int_0^t {\rm d}t_1\,\int_0^t
              {\rm d}t_2\,
              \langle\beta_\perp(t_1)\beta_\perp(t_2)\rangle
              \nonumber\\ &\,\simeq\,& \omega_{{\rm
                  L},0}^2{\beta_{\parallel,\rm i}^2\over
                \beta_{\perp,\rm i}^2}\int_0^t {\rm d}t_1\,\int_0^t
                          {\rm d}t_2\int_0^{t_1}{\rm
                            d}t_3\,\int_0^{t_2}{\rm
                            d}t_4\,\nonumber\\ & & \quad\quad
                          \beta_{x,0}(t_3)\beta_{x,0}(t_4)\langle
                          b_z(t_3)
                          b_z(t_4)\rangle\nonumber\\ &\,\simeq\,&\omega_{{\rm
                              L},0}^2{\beta_{\parallel,\rm i}^2\over
                            \beta_{\perp,\rm i}^2}\int_0^t{\rm
                            d}t_1\int_{t_1}^{t}{\rm
                            d}t_2\int_0^{t_1}{\rm d}t_3\, {A^2\over
                            3}\tau_{\rm
                            c}\,\beta_{x,0}^2(t_3)\nonumber\\ &\,\simeq\,&
                          {A^2\over 3}\omega_{{\rm L},0}^2\tau_{\rm
                            c}{\beta_{\parallel,\rm
                              i}^2\over\beta_{\perp,\rm i}^2}\,P(t)\ .
\end{eqnarray}
In the above equation, $\beta_{x,0}(t)$ represents the unperturbed
trajectory: $\beta_{x,0}\,=\,\beta_{x,\rm i}\cos(\omega_{{\rm
  L},0}t)-\beta_{z,\rm i}\sin(\omega_{{\rm L},0}t)$.  The function $P(t)$
contains powers of $t$ up to $t^3$, as well as sine and cosine
functions of $2\omega_{{\rm L},0}t$; its dimension is that of
$c^2\omega_{\rm L}^{-3}$. Noise will then dominate over the
unperturbed trajectory $x_{\perp,0}\,=\,\beta_{\perp,\rm i} ct$ over a
unperturbed return timescale if:
\begin{equation}
A^2\,\gtrsim\, {1\over \omega_{{\rm L},0}\tau_{\rm c}}\ ,
\end{equation}
or, equivalently:
\begin{equation}
A\,\gtrsim\, {r_{\rm L}\over \lambda_{\rm c}}\ .\label{eq:ineq-d}
\end{equation}
As before, $r_{\rm L}$ denotes the Larmor radius measured with respect
to the total magnetic field. This condition on $A$ is very similar to
that obtained upstream, since $r_{\rm L\vert u}\,\sim\, \Gamma_{\rm
  sh}^2 r_{\rm L\vert d}$ and $\lambda_{\rm c\vert
  u}\,\sim\,\Gamma_{\rm sh}\lambda_{\rm c\vert d}$, while $A_{\vert
  \rm u}\,\simeq\,A_{\vert\rm d}$.

According to the above analysis, and following Lemoine, Pelletier \&
Revenu (2006), phase mixing should be sufficiently large to erase all
dependence on the phase of the velocity vectors, hence Fermi
acceleration should be fully operative, if either one of the above
conditions on $A$ is satisfied. 

It is interesting to discuss the above results in light of recent
Monte Carlo numerical simulations of relativistic Fermi acceleration
in the presence of short scale turbulence (Niemiec, Ostrowski \& Pohl
2006). These authors have investigated the efficiency of the Fermi
process for various turbulence configurations, including a coherent
magnetic field, a long wave turbulence and a short scale
component. They confirm that Fermi acceleration is not operative if
$\delta B/B=0$, but find that a spectrum can develop over about two
orders of magnitude when $\delta B/B\,=\,80$ (see their Fig.~1). The
high energy break can be directly interpreted as the energy at which
the above inequality Eq.~(\ref{eq:ineq-d}) is no longer
fulfilled. Beyond that point, the spectrum steepends and the Fermi
process becomes inoperative. Interestingly, these authors also show
that if the pitch angle scattering amplitude in the short scale
turbulence is independent of energy, the break disappears and the
spectrum continues without bound. This is also expected, insofar as
the break that is determined by Eq.~(\ref{eq:ineq-d}) stems from the
fact that the scattering time in the short scale turbulence scales as
the square of the Larmor time whereas the unperturbed trajectory in
the background field depends linearly on the Larmor time. Assuming a
constant pitch angle scattering amplitude in the notations of Niemiec,
Pohl \& Ostrowski (2006) implies that the two scattering timescales
evolve in the same way, hence if short scale turbulence suffices to
isotropize directions at a given energy, it will do so at all
energies.

In order to produce powerlaw spectra over a large dynamic range, it is
thus necessary to reach an amplification factor $\delta B/B$ as large
as possible, since this ratio determines the dynamic range of particle
energies. Interestingly, the afterglow modeling of gamma-ray bursts
suggest that indeed, the usptream and downstream magnetic fields have
been amplified by a large factor.  In the range of energies in which
Fermi acceleration is operative, it is likely that the spectral index
would be equal to the so-called canonical value $s=2.3$, although the
generality of this prediction for different shapes of the $3$d
turbulence spectrum remains to be studied. The results of Niemiec,
Ostrowski \& Pohl (2006) cannot be used to infer this spectral index,
since they have restricted their analysis to values of $\delta B/B <
100$, and the spectral indices they report have been measured beyond
the break energy.  Techniques developed in Lemoine \& Pelletier
(2003), Lemoine \& Revenu (2006) are particularly suited to study this
problem and calculations are underway.

\section{Instabilities at perpendicular shock waves}\label{sec:instab}

The magnetic field in the shock front $B_{\vert\rm sh}$ is related to
the magnetic field in the upstream frame $B_{\vert\rm u}$ and the
associated electric field $E_{\vert\rm u}$ through the Lorentz
transform:
\begin{eqnarray}
B_{\parallel\vert\rm sh}&\,=\,& B_{\parallel\vert\rm
  u}\nonumber\\ \mathbf{B}_{\perp\vert\rm sh}&\,=\,& \Gamma_{\rm
  sh}\left(\mathbf{B}_{\perp\vert\rm u} - \mathbf{\beta}_{\rm
  sh}\times\mathbf{E}_{\perp\vert\rm u}\right)\ .
\end{eqnarray}
To zeroth order, the electric field in the upstream plasma frame
vanishes, hence the magnetic field in the shock front frame is mostly
perpendicular unless $\mathbf{B}_{\vert\rm u}$ is aligned along the
shock normal to within an angle $\sim {\cal O}(1/\Gamma_{\rm sh})$
(subluminal shock).  It then suffices to consider the fully
perpendicular situation with $B_{\parallel\vert\rm sh}=0$.

The cosmic rays stream ahead of the shock wave, carrying a net charge
density $\rho_{\rm cr}$ which will induce a counteracting charge
density $\rho_{\rm pl}$ in the background plasma. Note that the cosmic
rays do not induce an electrical current at zeroth order in the shock
front frame, only a net charge density. Since we consider the
generation of short scale turbulence, we neglect the cosmic ray
response to this short scale turbulence other than its effect on the
cosmic ray distribution scale $\ell_{\rm cr}$. For simplicity, we
approximate the cosmic-ray charge profile with a top-hat distribution.

As usual, we search for a stationary regime which serves as a basis
for perturbating the equations in the time-dependent regime. The
details of the calculations are provided in Appendix~\ref{sec:appperp}
for both stationary and time dependent quantities. In particular, the
stationary regime exhibits the following set-up:
\begin{equation}
\mathbf{B}=B_y \mathbf{e_y},\,\, \mathbf{u}=u_x\mathbf{e_x} +
u_z\mathbf{e_z} \ .
\end{equation}
Recalling that the shock normal is directed along $x$, $u_x\simeq
-\Gamma_{\rm sh}\beta_{\rm sh}$ characterizes the velocity of the
upstream inflowing through the front. The velocity component $u_z$
along the front is small compared to $u_x$, but its shear $\partial_x
u_z$ cannot be neglected (see Appendix~\ref{sec:appperp}).

\subsection{Linear analysis of the reduced system}

The complete system that governs the linear evolution is of seventh
order (see Appendix~\ref{sec:appperp}) and thus rather
involved. Nevertheless, it remains possible to derive the main results
since the Lorentz transform from the upstream comoving frame to the
shock front frame dominates the MHD propagation effects when the wave
scale is ``not too small", as will be made more precise in a next
subsection. In other words, the instability growth rates can be
derived to lowest order by assuming vanishing Alfv\'en and sound
velocities. We have verified that the analysis of the system including
$\beta_{\rm A}$ and $\beta_{\rm s}$ (but neglecting terms in $u_z$)
does not modify the growth rates obtained further below.

One should point out that electromagnetic waves with dispersion
relation $\omega_{\vert\rm u}(\mathbf{k_{\vert\rm u}})$ are Lorentz
transformed into waves with dispersion relation:
\begin{equation}
\omega\,\simeq\, \beta_{\rm sh}ck_x + {\cal
  O}\left(\frac{k_{y,z}}{\Gamma_{\rm sh}}\right)\ .
\end{equation}
In particular, for Alfv\'en waves, the next term on the r.h.s. is
$\beta_{\rm A}k_yc/\Gamma_{\rm sh}$. Note that the Alfv\'en velocity
is always expressed in the upstream plasma rest frame,
i.e. $\beta_{\rm A}\,=\,B_{y\vert \rm u}/\sqrt{4\pi\rho_{\rm
    u}c^2}$. Therefore it suffices in what follows to consider the
limit $k_z\rightarrow 0$. The limit $k_y\rightarrow 0$ can also be
considered but it appears more restrictive, because it limits the
analysis to the evolution of upstream magnetosonic modes. We will thus
consider both cases $k_y=0$ and $k_y$ finite. Of course, one can use
Fourier analysis without mode coupling as long as the spatial
dependence of the system coefficients can be neglected, in agreement
with our previous approximations, and in particular, that the cosmic
ray charge density is modeled as a step function.  In the shock front
rest frame, the physical picture of the instability is then as
follows. Impingent electromagnetic waves propagate in vacuum beyond
the length scale $\ell_{\rm cr}$ which characterizes the spatial
distribution of cosmic rays upstream of the shock front. For $0<x<
\ell_{\rm cr}$, the presence of the cosmic-ray charge density induces
a short scale instability. Hence we consider electromagnetic waves and
seek a solution in Fourier space with $\omega$ set by its vacuum
dispersion relation, but with a complex $k_x$ characterizing the
amplification in the charge layer $0<x< \ell_{\rm cr}$.

In this perpendicular configuration, the return current (in the
upstream frame) is not responsible for a supplementary tension effect,
but for a supplementary compression effect.  A $b_y$ perturbation
leads to a supplementary vertical (i.e. along $z$) compression that
can push like the kinetic compression. If we consider a spatial
modulation along the mean field ($k_y \neq 0$), a vertical
perturbation $b_z$ generates a supplementary compression in the
direction of the mean field that can push in phase with the kinetic
compression as well.

For large $\Gamma_{\rm sh}$ even with $k_z \neq 0$, the system reduces
to fourth order and can be expressed as the coupling between the
propagation of the vertical perturbed motion $\delta u_z$ and the
propagation of kinetic compression $\delta w/w$, where $w \equiv
\rho_{\rm u}c^2$ is the relativistic proper enthalpy density of the
cold upstream plasma.  The system of the two coupled equations can be
expressed using the system of
Eqs.~(\ref{eq:tdb}),(\ref{eq:tdw}),(\ref{eq:tdu}) in the limit
$\beta_{\rm A}\rightarrow 0$, $\beta_{\rm s}\rightarrow 0$. In
particular in this limit, one notes that $b_x\,\approx\,0$.

Then one obtains a single equation for $\delta u_z$ (see
Appendix~\ref{sec:perteq}) where derivatives $\partial_z$ cancel out
and where the derivative of $u_z$ is inserted, its second derivative
being neglected (we also assume $\beta_{\rm sh} = 1$ in the coefficients):
\begin{eqnarray}
\label{eq:maseq}
\hat D^4 \delta u_z - \left(\kappa^2 + u_z \kappa
\partial_x\right)\hat D^2 \delta u_z + \Gamma_{\rm sh}\kappa^2
\partial_x \hat D \delta u_z + \Gamma_{\rm sh}^2\kappa^2 \partial_y^2
\delta u_z = 0 \ .
\end{eqnarray}
The differential operator $\hat D$ is defined by $\hat{D} =
\Gamma_{\rm sh}\left(c^{-1}\partial_{\rm t} - \beta_{\rm sh}
\partial_x\right)$.  For a detailed understanding of the instabilities
one has to notice that the charge is not completely screened by the
plasma as a supplementary electric field is generated along the flow
(i.e. oriented towards $-x$): $E_x = u_zB_y/\Gamma_{\rm sh}$, along
with a sheared vertical motion $u_z$ such that $\partial_x u_z =
\kappa$ (see Appendix~\ref{sec:appperp}).  The instabilities stem from
this sheared motion.  The parameter $\kappa$, which carries the
dimension of a wavenumber, is defined in Eq.~(\ref{eq:kappa}). To our
present order of approximation, it can be expressed as:
\begin{equation}
\kappa \,\equiv \, \frac{\rho_{\rm cr}B_y}{\Gamma_{\rm sh} w} \ .
\label{eq:kappa_main}
\end{equation}
Note that both $\rho_{\rm cr}$ and $B_y$ are here evaluated in the
shock front frame. In the front frame, the parameter $\kappa$ appears
divided by $\Gamma_{\rm sh}$, hence we introduce $k_{*} \equiv
\kappa/\Gamma_{\rm sh}$.

The reduced form of the equation (see Appendix~\ref{sec:perteq})
clearly shows the ordering (by setting $\hat D = k_x \beta_{\rm sh}\Gamma_{\rm sh} \tilde
D$, $\delta u_z = \beta_{\rm sh}\Gamma_{\rm sh} \tilde u_z$):
\begin{equation}
\label{eq:nw1}
\tilde D^4 \tilde u_z -\left({k_*^2 \over k_x^2} + {u_z \over \Gamma_{\rm
    sh}}{k_*\partial_x \over k_x^2}\right)\tilde D^2 \tilde u_z + {k_*^2
  \partial_x \over k_x^3} \tilde D \tilde u_z + {k_*^2 \over k_x^4}
\partial_y^2 \tilde u_z = 0
\end{equation}
The inhomogeneity is contained in $\kappa^2$ that vanishes at large
distance from the shock. At infinity $\hat D \delta u_z = 0$, which
implies that for a mode having a specified $k_x$, the pulsation
$\omega = \beta_{\rm sh} k_xc$, as expected. In the precursor where
$\kappa^2 \neq 0$, we follow this mode characterized by its frequency
and look at the modification of its spatial behavior. As explained
above, we solve this equation by setting $c^{-1} \partial_t \mapsto
i\beta_{\rm sh} k_x$, $\partial_x \mapsto ik_x(1-\varepsilon)$, so
that $\hat D \mapsto ik_x \beta_{\rm sh}\Gamma_{\rm sh} \varepsilon$
($\tilde D \mapsto i \varepsilon$), where $\varepsilon$ is a complex
number, the imaginary part of which characterizes the growth rate.  It
is obtained by solving the algebraic equation:
\begin{equation}
\label{eq:nw2}
\varepsilon^4 +2{k_*^2 \over k_x^2} \varepsilon^2 - {k_*^2 \over
  k_x^2} \varepsilon -\frac{k_*^2k_y^2}{k_x^4} + i{u_z \over
  \Gamma_{\rm sh}}{k_* \over k_x}(1-\varepsilon)\varepsilon^2 = 0
\end{equation}
Actually, because $\varepsilon$ is small, the equation can be
simplified to:
\begin{equation}
\varepsilon^4 - {k_*^2 \over k_x^2} \varepsilon -\frac{k_*^2k_y^2}{k_x^4} = 0 \ .
\end{equation}
The contribution in $u_z$ introduces a small correction to the real
part of the roots (most $u_z$ contributions can be canceled out by a
frame transformation, but that correction cannot be canceled out
because it corresponds to the electric field component $E_x$).

\subsubsection{Analysis of the case $k_y=0$}
Let us first analyze the case $k_y=0$, or less restrictively for
$k_y^2 \ll \vert \varepsilon^2 \vert\,k_x^2$. This is a pure
magneto-sonic compression with $\mathbf{b}$ along the mean field.  One
gets the following three roots:
\begin{equation}
\varepsilon\, \simeq\, -\left(\frac{k_*}{k_x}\right)^{2/3} \left(-1, \,
\frac{1\pm i\sqrt{3}}{2}\right),
\end{equation}
One obtains an instability spatial growth rate $\gamma_x={\rm
  Im}\left(k_x\varepsilon\right)$ that increases with large $k_x$.
\begin{equation}
\gamma_x\, =\, \frac{\sqrt{3}}{2}k_*^{2/3}k_x^{1/3} \quad (k_x\, \gg\,
k_*) \ .\label{eq:g2}
\end{equation}
Thus the characteristic wave number associated with the charge in this
problem of magneto-sonic wave propagation is $k_* \equiv
\kappa/\Gamma_{\rm sh}$.

\subsubsection{Analysis of the case $k_y \neq 0$}
With $k_y \neq 0$, one recovers the previous growth rate if $k_y \ll
k_*^{1/3}k_x^{2/3}$, while for $k_y \gg k_*^{1/3}k_x^{2/3}$ one
derives :
\begin{equation}
\varepsilon\, \simeq\, \left(\frac{k_y k_*}{k_x^2
}\right)^{1/2}(1,-1,i,-i)\ ,\label{eq:g4}
\end{equation}
or, equivalently:
\begin{equation}
\gamma_x \,\simeq\, \left(k_y k_*\right)^{1/2} \quad
\left(k_y\,\gg\,k_*^{1/3}k_x^{2/3}\right)\ .\label{eq:g5}
\end{equation}

The spatial growth rates are thus $\gamma_x \sim (k_*^2k_x)^{1/3}$ and
$(k_*k_y)^{1/2}$ respectively. Remarkably they are independent of
$k_z$ that can be chosen arbitrarily, provided it remains small
compared to $k_x$ for the consistency of the derivation.

\subsubsection{Comparison to MHD and cosmic-ray length scales}
One should compare the scale defined by $k_*$ to the smallest scale
$l_{\rm MHD}$ for our MHD description. Note that this length scale is
defined in the comoving upstream frame. Since $k_{x\vert\rm
  u}\,\simeq\,k_x/\Gamma_{\rm sh}$ while $k_{y\vert\rm u}\,\simeq\,
k_y$, it suffices to require $k_* l_{\rm MHD}\,<\,1$ to ensure that
there exist modes in the MHD range with $k>k_*$.  One obtains:
\begin{equation}
k_* l_{\rm MHD}\, =\,\frac{n_{\rm cr}eB_y}{\Gamma_{\rm
    sh}^2w}\beta_{\rm A}r_{0\vert\rm u}
\end{equation}
which can be rewritten in terms of the fraction $\xi_{\rm cr}$ of
(downstream) shock internal energy converted into (downstream) cosmic
ray energy:
\begin{equation}
\xi_{\rm cr} \,\equiv\, \frac{e_{\rm cr\vert d}}{4\Gamma_{\rm sh}^2w} \ .
\end{equation}
To this effect, we assume that the accelerated population can be
described as a power-law of index $-s$ and minimum momentum $p_{\rm
  min}$, so that:
\begin{equation}
n_{\rm cr}\,\simeq\, \frac{\vert 1-s\vert}{\vert 2-s\vert}\frac{e_{\rm
    cr}}{p_{\rm min}c}\ .
\end{equation}
This equation assumes $s>2$; if $s=2$, then $\vert 2-s\vert$ should be
replaced by $\log(p_{\rm max}/p_{\rm min})$. Since $e_{\rm cr\vert
  d}\,\sim\,e_{\rm cr\vert sh}$ (as $\Gamma_{\rm sh\vert
  d}=\sqrt{9/8}$ for a strong ultra-relativistic shock), one obtains:
\begin{equation}
k_* l_{\rm MHD} \,\simeq\, 4\frac{\vert 1-s\vert}{\vert
  2-s\vert}\xi_{\rm cr}\beta_{\rm A} \frac{r_{0\vert\rm u}}{r_{\rm
    min\vert\rm u}}\Gamma_{\rm sh}^2 \ .\label{eq:totgrow}
\end{equation}
Here $r_{\rm min\vert u}$ denotes the minimum Larmor radius of the
accelerated population, as measured upstream. The typical energy of
the first generation of cosmic rays is $\Gamma_{\rm sh}^2 m_{\rm
  p}c^2$, so that $r_{\rm min\vert u} \sim \Gamma_{\rm sh}^2
r_{0\vert\rm u}$. Therefore
\begin{equation}
k_* l_{\rm MHD} \,\sim\, 4\frac{\vert 1-s\vert}{\vert
  2-s\vert}\xi_{\rm cr} \beta_{\rm A} \,\ll\, 1 \ .
\end{equation}
Thus $k_*$ defines an MHD scale.

In order for the instability to be efficient, one must also require
$\gamma_x \ell_{\rm cr}\,\gg\,1$, which is a non-trivial requirement in
view of the restricted value of $\ell_{\rm cr}$. One finds:
\begin{equation}
\gamma_x \ell_{\rm cr}\,\simeq\, 4\frac{\vert 1-s\vert}{\vert
  2-s\vert}\frac{\gamma_x}{k_*}\xi_{\rm cr}\, g\ ,\label{eq:gro}
\end{equation}
with $g\,\equiv\,1$ for large scale turbulence, and
$g\,\equiv\,r_{0\vert\rm u}/\lambda_{\rm c\vert u}$ for short scale
turbulence. In the latter case, one may assume $\lambda_{\rm c\vert
  u}\,\simeq\, l_{\rm MHD}$, as suggested by the fact that the growth
rate increases with $k$, so that $g\,\sim\,1/\beta_{\rm A}$. Then, in
both cases, the total growth is presumably much larger than unity if
$\xi_{\rm cr}$ is not too small and $k_x$, $k_y$ sufficiently large as
compared to $k_*$.  Hence even the first generation of cosmic rays is
able to destabilize the upstream magnetic field on the shortest
scales. The effect of the high energy part of the accelerated
population will be addressed shortly.

As mentioned, the study of the more extended system including the
contribution of the terms involving the Alfv\'enic and sonic
contributions (but neglecting terms of order $u_z$ in the system)
confirms the results above for the growth rates, provided $\beta_{\rm
  A}\,\ll\,1$ and $\beta_{\rm s}\,\ll\,1$.

\subsubsection{Magneto-sonic saturation}
One can estimate the amplitude of the various components in terms of
$b_y$. It has already been mentioned that $\vert b_x \vert \ll \vert
b_y \vert$,
\begin{equation}
b_y \,\simeq\, \frac{\beta_{\rm sh}}{\Gamma_{\rm sh}} \delta u_x
\end{equation}
and
\begin{equation}
b_z \,\simeq\, -\frac{1}{\kappa}\partial_y \delta u_x \ .
\end{equation}
Thus we find that $b_z$ may achieve a large amplitude, because
\begin{equation}
b_z \,\simeq\, - \frac{\Gamma_{\rm sh}}{\kappa \beta_{\rm sh}} \partial_y b_y \ .
\end{equation}
In the frame of linear theory, the saturation level of these
magneto-sonic instabilities is simply estimated by the fact that the
compression has a limited amplitude $\vert \delta w/w \vert
<1$. Defining the power spectrum of $b_y$ and $b_z$ per log interval
of wavenumber ${\cal P}_{b_y}\,\equiv\, (k^3/2\pi^2)\vert\tilde
b_y\vert^2$ in terms of the Fourier component $\tilde b_y$ and
similarly for ${\cal P}_{b_z}$ in terms of $\tilde b_z$, one derives
from Eq.~(\ref{eq:redsys}):
\begin{equation}
{\cal P}_{b_y}^{1/2} \,\leq\, \frac{k_*^2}{\vert \varepsilon^2k_x^2
  + 2k_*^2 \vert}
\end{equation}
and
\begin{equation}
{\cal P}_{b_z}^{1/2}\,=\, \frac{k_y}{k_*} {\cal P}_{b_y}^{1/2} \ .
\end{equation}
Note that the power spectrum is normalized to the ratio of amplitude
of the turbulent (small scale) component to the coherent $B_0$.
Consider the first branch of the instability, given by
Eq.~(\ref{eq:g2}) which applies in the limit $k_x\,\gg\,k_*$,
$k_y\,\ll\,k_*^{1/3}k_x^{2/3}$. There one can show that ${\cal
  P}_{b_y}^{1/2}$ saturates at a value of order
$\left(k_x/k_*\right)^{-2/3}\,\ll\,1$, but ${\cal P}_{b_z}^{1/2}$
saturates at a value of order unity when
$k_y\,\sim\,k_*^{1/3}k_x^{2/3}$. For the second branch, given in
Eq.~(\ref{eq:g5}) and which applies in the limit
$k_y\,\gg\,k_*^{1/3}k_x^{2/3}$, one finds that ${\cal P}_{b_y}^{1/2}$
saturates at a value of order $k_*/k_y\,\ll\,1$, but again ${\cal
  P}_{b_z}^{1/2}$ saturates at a value unity. This moderate saturation
level can be directly traced back to the saturation of these
compressive modes at the linear level $\vert\delta w/w\vert\sim 1$. As
discussed in Sec.~\ref{sec:fermi}, such a level $\delta B/B_0\sim1$
should not allow a powerlaw spectrum to develop through Fermi
acceleration.

Nevertheless, our present analytical description is by definition
limited to the linear regime. It would certainly be interesting to
pursue the calculations using numerical simulations as one could
expect that the holes produced in the plasma would widen with the
growing magnetic pressure inside and that the density excesses become
spikes. Thus one might reasonably expect the generation of many local
small magneto-sonic shocks that are absorbed by the main
ultra-relativistic shocks, with the reconversion of some amount of
cosmic-ray energy into thermal energy in the precursor.

The previous discussion has focused on the role played by the first
generation of cosmic rays, since higher energy cosmic rays can only be
present if acceleration to smaller energies has been
completed. However it is easy to verify that, if highest energy cosmic
rays are present in the shock precursor, they are bound to play a
dominant role in the amplification of the magnetic field, most notably
because the length scale of the cosmic ray distribution increases as
$r_{\rm L}^2$. For instance, at a distance $x$ from the shock front
where only cosmic rays of momentum larger than $p_*$ can be found, the
critical wavenumber for the instability reads for $s > 2$ ($s$
spectral index of the accelerated population):
\begin{equation}
k_*\,=\, k_{*,0} \left(\frac{p_*}{p_{*,0}}\right)^{1-s}\ ,
\end{equation}
where $k_{*,0}$ denotes the same wavenumber evaluated for the whole
cosmic-ray distribution, and $p_{*,0}$ denotes the minimum momentum of
the cosmic ray distribution at the shock front. Consequently, if
$\gamma_x\propto k_*^a$, with $a=2/3$ or $1/2$ for the two branches of
the instability obtained above:
\begin{equation}
\gamma_x \ell_{\rm cr,*}\,=\,\left(\frac{p_*}{p_{*,0}}\right)^{b+a(1-s)}
\gamma_{x,0}\ell_{\rm cr}\ ,
\end{equation}
with $b=1$ for large scale turbulence, $b=2$ for short scale
turbulence.  Here as well, $\gamma_{x,0}$ should be understood as the
growth rate derived previously for the whole cosmic-ray
distribution. This product grows with increasing $p_*/p_{*,0}$
provided $s<(b+a)/a$, i.e. $s<8$ ($b=2$, $a=2/3$), or $s<5$ ($b=2$,
$a=1/2$), or $s<5/2$ ($b=1$, $a=2/3$) or finally $s<3$ ($b=1$,
$a=1/2$). These inequalities are likely to be fulfilled.  Finally, the
product $k_* l_{\rm MHD}$ scales as $(p_*/p_{*,0})^{1-s}$ hence
decreases if $s>1$. This also means the typical scale of the
instability increases with increasing $p_*/p_{*,0}$.

Of course, the above implicitly ignores the spatial dependence of $B$,
but it suggests that amplification through the streaming of the
highest energy part of the cosmic ray distribution will seed much more
efficiently the magnetic field amplification.

\subsubsection{Non-existence of an incompressible instability}

Since the above compressive instabilities appear to saturate at a
level too small ($\delta B/B\sim 1$) to allow Fermi acceleration over
a broad range of energies, the possible existence of non-compressive
instability becomes crucial. One can search for such modes by
selecting the wave vectors accordingly, in which case the reduced
system becomes:
\begin{eqnarray}
\tilde D^2 \tilde u_z -\left(\frac{k_*^2}{k_x^2}+\frac{k_*
  \partial_z}{k_x^2} + \frac{u_z}{\beta_{\rm sh}\Gamma_{\rm sh}}\frac{k_*
  \partial_x}{k_x^2}\right)\tilde u_z + \frac{k_*}{k_x}
& = & 0 \nonumber \\ \frac{k_*\partial_x}{k_x^2} \tilde
D \tilde u_z - \frac{k_*\partial_y^2}{k_x^3} \tilde u_z
- \frac{\partial_z}{k_x} \tilde D^2 \tilde u_z & = & 0
\end{eqnarray}
It is easy to verify that this system possesses no unstable solution
but only damped solutions. Hence incompressible unstable modes do not
exist in this MHD description.

\section{Conclusions}\label{sec:conc}
In this study, we have examined the amplification of a pre-existing
magnetic field upstream of an ultra-relativistic shock wave. We have
assumed that the magnetic field is fully perpendicular to the shock
normal in the shock front frame, as generally expected in the
ultra-relativistic limit $\Gamma_{\rm sh}\,\gg\,1$. In the shock
frame, the cosmic rays do not induce any current at zeroth order, only
a net charge density, which is partly screened by the inflowing
plasma. This charge distribution then triggers an instability on very
short spatial scales, with a growth rate increasing with the
wavenumber. Cosmic rays do not respond to this amplification as it
takes on scales much shorter than the typical Larmor radius.

It is important to stress that these instabilities are of a different
nature than the resonant or non-resonant instability studied by Bell
(2004) in the case of a non-relativistic parallel shock wave, most
notably because they are essentially {\it compressive}. Since density
depletions are naturally limited, our linear analysis indicates that
these instabilities saturate at a moderate level of amplification,
$\delta B/B_0 \, \sim\, 1$. Furthermore, there is no unstable
incompressible mode within the present MHD approximation which could
provide a higher level of amplification. Therefore, we conclude that,
within the framework of ideal MHD, instabilities at ultra-relativistic
magnetized shock waves appear limited in their efficiency.

Such instabilities cannot therefore account for the degree of
amplification that has been inferred from the modeling of the
afterglow radiation of gamma-ray bursts. We have also argued in
Section~\ref{sec:fermi} that the ratio $\delta B/B_0$ sets the dynamic
range of energies over which powerlaw spectra can be produced through
Fermi acceleration. The present instabilities thus cannot allow
successful Fermi acceleration. Nevertheless, it would certainly be
useful to pursue the present investigation with dedicated numerical
simlations, which would allow one to go beyond the present linear
approximation and study whether non-linear effects can push the small
scale magnetic field to higher values.

If the standard interpretation of gamma-ray burst afterglows as the
synchrotron light of electrons accelerated at the forward shock (but
see Uhm \& Beloborodov 2006, Genet, Daigne \& Mochkovitch 2007 for
recent alternatives involving the reverse shock) holds, one is led to
conclude that some other instability is able to amplify the upstream
magnetic field on short spatial scales, or that some other form of
acceleration mechanism is operating (see for instance Hoshino
2008 for an alternative involving radiative pressure effects). 
The present work indicates that such cosmic ray induced instabilities
would
involve non-MHD effects. The short spatial scales $\lambda_{\rm
  c}\,\ll\,r_{\rm L}/\Gamma_{\rm sh}$ required are in conflict with
the usual synchrotron resonance between cosmic rays and MHD waves;
this remark is actually one of the motivation of the present work in a
full MHD framework. However, one cannot rule out that other types of
resonance could become relevant and yet involve incompressible modes,
such as Alfv\'en waves. One promising instability, currently under
study, involves a Cerenkov resonance ($\omega = k_xc$) between plasma
waves and the cosmic ray beam, generating a modified two stream
instability. The growth rates appear quite promising and the details
will be given in a forthcoming article.

\appendix

\section{Background and perturbed quantities for perpendicular shock
  waves}\label{sec:appperp}

\subsection{Stationary regime}

In the shock front frame, the electric field, current and charge
density read:
\begin{eqnarray}
\mathbf{E}&\,=\,&-\frac{1}{\gamma}\mathbf{u}\times\mathbf{B}\,=\,
\frac{u_zB_y}{\gamma}\mathbf{e_x}-\frac{u_xB_y}{\gamma}\mathbf{e_z}\ ,\nonumber\\
\mathbf{j}&\,=\,&\frac{c}{4\pi}\partial_xB_y\mathbf{e_z}\ ,\nonumber\\
\rho_{\rm pl}&\,=\,&-\rho_{\rm cr} + \frac{1}{4\pi}\partial_xE_x\ .
\end{eqnarray}
The four velocity of the inflowing plasma is written $(\gamma
c,\mathbf{u}c)$ (hence $\mathbf{u}=\gamma\mathbf{\beta}$ is
dimensionless). In this equation, $\gamma$ represents the total
Lorentz factor of the upstream plasma; it will be shown to coincide
nearly with $\Gamma_{\rm sh}$ in the following.

Note that in the shock front, there is no current at zeroth order
associated with the cosmic rays, only a net charge density $\rho_{\rm
  cr}$ which may be partially screened by an induced non-zero plasma
charge density $\rho_{\rm pl}$. Faraday's law implies in this
stationary regime:
\begin{equation}
\mathbf{\nabla}\times\mathbf{E}=0\,\leftrightarrow\,\partial_x
\left(\frac{u_xB_y}{\gamma}\right)\,=\,0\ ,\label{eq:Flaw}
\end{equation}
Hence the Gauss equation can be rewritten as:
\begin{equation}
\rho_{\rm pl}\,=\,-\rho_{\rm cr} +
\frac{B_y}{4\pi\gamma}\left(\partial_xu_z - \frac{u_z}{u_x}\partial_x
u_x\right)\ .\label{eq:Glaw}
\end{equation}

Energy momentum conservation for the upstream fluid implies:
\begin{equation}
\rho_{\rm u}c^2\mathbf{u}\cdot\mathbf{\nabla}\mathbf{u}\,=\,\rho_{\rm
  pl}\mathbf{E} +\frac{1}{c}\mathbf{j}\times\mathbf{B}\ ,
\end{equation}
where $\rho_{\rm u}$ is the proper mass density of the upstream
plasma. Denoting $F_{\rm m}\equiv -\rho_{\rm u}c^2 u_x$ the mass flux
(times $c$) through the shock front:
\begin{eqnarray}
F_{\rm m}\partial_xu_x&\,=\,&\left[\rho_{\rm cr} -
  \frac{B_y}{4\pi\gamma}\left(\partial_xu_z - \frac{u_z}{
    u_x}\partial_xu_x\right)\right]\frac{u_zB_y}{\gamma}\,\nonumber\\ &&\quad\,+\,
\frac{1}{4\pi}B_y\partial_xB_y\ ,\nonumber\\ F_{\rm
  m}\partial_xu_z&\,=\,&-\left[\rho_{\rm cr} - \frac{B_y}{
    4\pi\gamma}\left(\partial_xu_z - \frac{u_z}{
    u_x}\partial_xu_x\right)\right]\frac{u_xB_y}{\gamma}\ .\label{eq:mom1}
\end{eqnarray}

Since: $\gamma=(1 + u_x^2 + u_z^2)^{1/2}$, 
\begin{equation}
\partial_x\gamma = \frac{u_x\partial_xu_x + u_z\partial_xu_z}{\gamma}\
,
\end{equation}
therefore (\ref{eq:Flaw}) implies:
\begin{equation}
\partial_xB_y\,=\,-B_y \frac{\gamma}{u_x}\left(
	\frac{\gamma^2-u_x^2}{\gamma^3}\partial_xu_x -
	\frac{u_xu_z}{\gamma^3}\partial_xu_z\right)\ . \label{eq:dB}
\end{equation}

Using (\ref{eq:dB}), the system (\ref{eq:mom1}) can be transformed
into:
\begin{eqnarray}
\left(F_{\rm m} - \frac{B_y^2}{4\pi\gamma^2u_x}\right)\partial_xu_x
&\,=\,& \rho_{\rm cr}\frac{u_zB_y}{\gamma}\, \nonumber\\ \frac{B_y^2
  u_z}{4\pi\gamma^2}\partial_xu_x + \left(F_{\rm m} -
\frac{B_y^2u_x}{4\pi\gamma^2}\right)\partial_xu_z&\,=\,&-\rho_{\rm
  cr}\frac{u_xB_y}{\gamma}\ .\label{eq:mom2}
\end{eqnarray}

This sytem can be solved by successive approximations. For instance
the ratio:
\begin{equation}
\left\vert\frac{B_y^2}{4\pi\gamma^2u_xF_{\rm m}}\right\vert \sim \frac{1}{
  \Gamma_{\rm sh}^2}\beta_{\rm A}^2\ll 1\,
\end{equation}
since $\vert u_x\vert\sim\Gamma_{\rm sh}$; $\beta_{\rm A}=B_{y|u}/\sqrt{4\pi \rho_u c^2}$ 
represents the Alfv\'en velocity (as measured in the upstream rest frame). Hence:
\begin{equation}
\partial_xu_x\,\simeq\,\rho_{\rm cr}\frac{u_zB_y}{ \gamma F_{\rm m}}\ .
\end{equation}
Consider now the ratio of the first term on the lhs of the second
equation in (\ref{eq:mom2}) to the term of the rhs of the same
equation:
\begin{equation}
\left\vert\frac{\partial_xu_x \frac{B_y^2u_z}{4\pi\gamma^2}}{
\rho_{\rm cr}\frac{u_xB_y}{\gamma}}\right\vert\,\sim\,\frac{u_z^2}{
  \Gamma_{\rm sh}^2}\beta_{\rm A}^2 .
\end{equation}
This ratio is much smaller than unity (since $u_z\,\ll\,\Gamma_{\rm
  sh}$). Then:
\begin{eqnarray}
\partial_xu_z&\,\simeq\,&\rho_{\rm cr}\frac{B_y}{F_{\rm m} +
  \frac{B_y^2}{4\pi\gamma}}\,\nonumber\\
&\,\Rightarrow\, & u_z(x)\,\simeq\, \frac{B_y}{ F_{\rm m} +
  \frac{B_y^2}{4\pi\Gamma_{\rm sh}}}\int_{x*}^{x}\rho_{\rm cr}\mathrm{d}x\ .
\end{eqnarray}
In the last equation, $x_*$ corresponds to the maximal distance to
which cosmic rays can stream ahead of the shock front, as measured in
the shock front frame. We also assume that $u_x$, $B_y$ and $\rho$
vary on much longer scales than $u_z$, and that $\vert
u_z\vert\ll\vert u_x\vert$. This hierarchy can be verified by taking
the ratios of the equations in (\ref{eq:mom1}).

Since $B_y^2/(4\pi\gamma)\,\approx\,\beta_{\rm A}^2 F_{\rm m}$, one
can simplify the previous expression and obtain the following order of
magnitude:
\begin{equation}
\vert u_z\vert\,\sim\,\frac{\ell_{\rm cr}}{r_{\rm L *}}\frac{e_{\rm cr}}{
  \Gamma_{\rm sh}\rho_{\rm u}c^2}\ .
\end{equation}
To obtain this result, we have approximated the integral over
$\rho_{\rm cr}$ as $\ell_{\rm cr}\rho_{\rm cr}$, $\ell_{\rm cr}$
representing the length scale of the distribution in the shock frame,
and written $\rho_{\rm cr}\simeq e_{\rm cr}/(p_*c)$; $e_{\rm cr}$
denotes the cosmic ray energy density while $p_*$ ($r_{\rm L *}$)
represents their typical momentum (Larmor radius). Finally, $e_{\rm
  cr}/(\Gamma_{\rm sh}\rho_{\rm u}c^2)\,\sim\,\Gamma_{\rm sh}\xi_{\rm
  cr}$ where $\xi_{\rm cr}$ is the fraction of shock internal energy
carried away by the accelerated population.  In a straightforward way,
one obtains:
\begin{equation}
\frac{\ell_{\rm cr}}{r_{\rm L*}}\,\sim\,{\cal O}(1)\ ,
\end{equation}
so that:
\begin{equation}
\frac{\vert u_z\vert}{\vert u_x\vert}\,\sim\,\xi_{\rm cr}\,\ll\,1\ .
\end{equation}
This justifies the previous approximations.

\subsection{Time dependent quantities}

\subsubsection{Perturbed electric field}
The perturbed Ohm law leads to:
\begin{eqnarray}
\label{Eq:CE}
 \mathbf{\delta E} &\,=\, & \frac{1}{ \gamma} \biggl[ B_y \delta u_z +
u_z \delta B_y + \,\nonumber\\&&\,\,\,\,+\,\frac{1}{ \gamma^2}
\left(\Gamma_{\rm sh} \beta_{\rm sh} u_z \delta u_x B_y-\delta u_z
u_{\rm z}^2 B_y\right)\biggr ] \mathbf{e_x} + \nonumber \\ & &\frac{1}{
\gamma} \left(-\delta B_x u_z+\delta B_z u_{\rm x}\right) \mathbf{e_y}
+ \nonumber \\ & &\frac{1}{ \gamma} \biggl[\Gamma_{\rm sh} \beta_{\rm
sh} \delta B_y-\delta u_{\rm x} B_y + \,\nonumber\\&&\,\,\,\,+\,\frac{1}{
\gamma^2}\left(\Gamma_{\rm sh}^2 \beta_{\rm sh}^2 \delta u_x B_y
-\Gamma_{\rm sh} \beta_{\rm sh} \delta u_z u_z B_y\right)\biggr]
\mathbf{e_z} \ .
\end{eqnarray}

We keep track of the terms in $u_z$ in these equations in order to
properly evaluate the terms which involve the derivative of $u_z$ with
respect to $x$. This latter quantity is of the same order than other
background quantities and cannot be neglected. However, terms in $u_z$
will be neglected in the final equations of evolution of the various
background quantities since $u_z\ll u_x$. For similar reasons,
derivatives of background quantities (other than $u_z$) such as
$\rho_{\rm pl}$, $B_y$ and $u_x$ (hence $\gamma$) can be
neglected. Thus, setting $\gamma \simeq \Gamma_{\rm sh}$, $\beta
\simeq \beta_{\rm sh}$ we obtain:
\begin{eqnarray}
\label{Eq:CE2}
\mathbf{\delta E} &\,=\, & \biggl(\frac{1}{ \Gamma_{\rm sh}} B_y
\delta u_{\rm z} + \frac{1}{ \Gamma_{\rm sh}}u_z \delta B_y +
\frac{\beta_{\rm sh} }{ \Gamma_{\rm sh}^2} u_z B_y \delta u_{\rm
  x}\,\nonumber\\&&\,\,\,\,+\,- \frac{1}{ \Gamma_{\rm sh}^3} u_z^2 B_y
\delta u_z \biggr) \mathbf{e_x} + \nonumber \\ & & \left(-\beta_{\rm
  sh}\delta B_z -\frac{1}{ \Gamma_{\rm sh}} \delta B_x u_z\right)
\mathbf{e_y} + \nonumber \\ & & \biggl(\beta_{\rm sh} \delta B_{\rm
  y}- \frac{1}{ \Gamma_{\rm sh}^2}B_y \beta_{\rm sh} \delta u_z u_z -
\frac{1}{ \Gamma_{\rm sh}^3} B_y \delta u_x \biggr)\mathbf{e_z} \ .
\end{eqnarray}

\subsubsection{Perturbed equations}\label{sec:perteq}

The perturbed equations for the perturbed components of the magnetic
field, defined in units of $B_y$ as $\mathbf{b}=\mathbf{\delta
  B}/B_y$ can be written:
\begin{eqnarray}
\hat{D} b_x &\, =\, & \frac{1}{ \Gamma_{\rm sh}^2} \partial_y \delta
u_x + u_z \left(-\partial_z b_{\rm x}+\frac{\beta_{\rm sh} }{
\Gamma_{\rm sh}}\partial_y \delta u_z\right) \ , \nonumber \\ \hat{D}
b_y &\, =\, & -\partial_z \delta u_z - \frac{1}{ \Gamma_{\rm sh}^2}
\partial_x \delta u_x - \frac{\beta_{\rm sh}}{ \Gamma_{\rm sh}}u'_z
\delta u_z+ \nonumber\\ & & +u_z \biggl(-\partial_z b_y - \frac{\beta_{\rm
sh}}{ \Gamma_{\rm sh}}\partial_z \delta u_x - \frac{\beta_{\rm sh}
}{\Gamma_{\rm sh}} \partial_x \delta u_z + \nonumber \\ &
&\,\,\,\,+\frac{1}{ \Gamma_{\rm sh}^2} u_z\partial_z \delta u_z
\biggr)\ , \nonumber \\ \hat{D} b_z &\, =\, & \partial_y \delta u_{\rm
z} + u'_z b_x + u_z \biggl(\partial_x b_x + \partial_y b_y
+\frac{\beta_{\rm sh}}{ \Gamma_{\rm sh}} \partial_y \delta u_x+
\nonumber \\ & &\,\,\,\,- \frac{u_z}{ \Gamma_{\rm sh}^2} \partial_y
\delta u_z\biggr) \ .\label{eq:tdb}
\end{eqnarray}

The continuity equation can be expressed as:
\begin{equation}
\hat{D} \frac{\delta w}{ w} = - {\mathbf \nabla}\cdot{\mathbf \delta
  u} -{1\over c}\partial_t \delta \gamma \simeq -(\partial_x \delta
u_x+\partial_y \delta u_y+\partial_z \delta u_z) \ .\label{eq:tdw}
\end{equation}
We have noted $w = \rho_u c^2$ the proper enthalpy density of the
upstream plasma. In the perturbation regime the contribution of the
temporal derivative of the fluctuating part of the Lorentz factor is
negligible. Again, while deriving Eq.(\ref{eq:tdw}) the $x$
derivatives of zeroth order quantities have been discarded.

The equation of motion in the shock rest frame accounting for the
displacement current are:
\begin{equation}
\label{Eq:Mot}
w \hat{D} {\mathbf u}= \rho_{\rm Pl} {\mathbf E} - {\mathbf \nabla}
\frac{B^2}{ 8\pi} + {\mathbf B}\cdot{\mathbf \nabla}\, \frac{{\mathbf B}
  }{ 4\pi} -\frac{1}{4\pi c} \frac{\partial {\mathbf E}}{ \partial
  t} \times {\mathbf B} \ .
\end{equation}
Using Eq.(\ref{Eq:CE2}) and developping the Eq.(\ref{Eq:Mot}) to the
first order we finally get after some manipulations:
\begin{eqnarray}
\hat{D} \delta u_x & = & \beta_{\rm A}^2\Gamma_{\rm sh}^2
\left(\partial_y b_x -\partial_x b_y\right) -\kappa \delta u_z
+\nonumber\\ & & + \beta_{\rm A}^2\Gamma_{\rm sh}^2\left(\beta_{\rm
  sh} \frac{1}{c}\partial_{\rm t} b_y -\frac{1}{ \Gamma_{\rm sh}^3
  c}\partial_{\rm t} \delta u_x\right) - \beta_{\rm s}^2 \partial_x
\frac{\delta w}{ w} + \nonumber \\ & & + u_z\Biggl[\kappa \left(\beta_{\rm
    A}^2 -1\right)b_y + \frac{\beta_{\rm sh}}{\Gamma_{\rm
      sh}}\kappa(\beta_{\rm A}^2 -1)\delta u_x + \nonumber \\ & &
  \quad + \beta_{\rm sh}\Gamma_{\rm sh}\beta_{\rm A}^2\left(\partial_z
  b_y - \partial_y b_z\right) +\,\nonumber\\ & &\quad
  +\left(\beta_{\rm sh}^2\beta_{\rm A}^2 - \beta_{\rm A}^2
  -1\right)\partial_z\delta u_x + \nonumber \\ & &\quad +\beta_{\rm
    A}^2\left(\partial_x\delta u_z - \beta_{\rm
    sh}\frac{1}{c}\partial_t\delta
  u_z\right)\Biggr]\ ,\nonumber\\ \hat{D} \delta u_y & = & \beta_{\rm
  sh}\Gamma_{\rm sh}\kappa\, b_z -\beta_{\rm s}^2 \partial_y
\frac{\delta w}{w} + u_z\left(\kappa b_x - \partial_z\delta
u_y\right)\ , \nonumber \\ \left(1 +\beta_{\rm A}^2\right)\hat{D}
\delta u_z & = & \beta_{\rm A}^2\left(\partial_y b_z -\partial_z
b_y\right) - \beta_{\rm sh}\Gamma_{\rm sh} \kappa\left(1-\beta_{\rm
  A}^2\right) b_{\rm y}+ \nonumber \\ & & -\kappa\beta_{\rm
  sh}^2\left(1-\beta_{\rm A}^2\right) \delta u_x - \frac{\beta_{\rm
    sh}}{\Gamma_{\rm sh}}\beta_{\rm A}^2 \partial_z \delta u_x +
\nonumber \\ & & - \beta_{\rm s}^2 \partial_z \frac{\delta w}{ w} +
\beta_{\rm sh}\Gamma_{\rm sh}\kappa\frac{\delta w}{ w}+ \nonumber \\ &
& + u_z\Biggl[\frac{\beta_{\rm sh}}{\Gamma_{\rm
      sh}}\left(1-2\beta_{\rm A}^2\right)\kappa\delta u_z -
  \left(1+\beta_{\rm sh}^2\beta_{\rm A}^2\right)\partial_z\delta u_z +
  \nonumber \\ & &\quad + \beta_{\rm sh}\Gamma_{\rm sh}\beta_{\rm
    A}^2\left(\partial_y b_x - \partial_x b_y\right) + \nonumber \\ &
  &\quad+\beta_{\rm sh}\beta_{\rm A}^2\left(\frac{1}{c}\partial_t
  \delta u_x-\beta_{\rm sh}\partial_x\delta u_x\right) + \nonumber
  \\ & &\quad +\beta_{\rm A}^2\Gamma_{\rm
    sh}\left(\frac{1}{c}\partial_t b_y - \beta_{\rm sh}\partial_x
  b_y\right)\Biggr] \ .\label{eq:tdu}
\end{eqnarray}
We used the shorthand notation $\beta_{\rm s}^2=c_{\rm s}^2/c^2$ with
$c_{\rm s}$ the speed of sound, and:
\begin{equation}
\beta_{\rm A}^2 \,\equiv\, \frac{B_y^2}{ 4\pi \Gamma_{\rm sh}^2 w}\
,\label{eq:betaA}
\end{equation}
which defines the Alfv\'en velocity squared (as measured
upstream). The quantity $\kappa$, which carries the dimension of a
wavenumber, is defined as follows:
\begin{equation}
\kappa \,\equiv\, u_z'\,=\, \frac{1}{ 1+\beta_{\rm A}^2}\frac{\rho_{\rm
cr}B_y}{ \Gamma_{\rm sh}w}\ .\label{eq:kappa}
\end{equation}
The plasma charge density $\rho_{\rm pl}$ is indeed related to the
cosmic-ray charge density $\rho_{\rm cr}$ through the Maxwell
equation:
\begin{equation}
\rho_{\rm pl}\,=\,-\rho_{\rm cr} \,+\, \frac{B_y}{4\pi\Gamma_{\rm
    sh}}u_z'\ ,
\end{equation}
hence:
\begin{equation}
\frac{\rho_{\rm pl}B_y}{\Gamma_{\rm sh}w}\,=\,-\kappa\ .
\end{equation}

In the limit $\beta_{\rm A}\ll1$ and $\beta_{\rm s}\ll1$, the system
governing the evolution of the perturbed quantities reads:
\begin{eqnarray}
\hat{D} b_x &\, =\, & \frac{1}{ \Gamma_{\rm sh}^2} \partial_y \delta
u_x + u_z \left(-\partial_z b_{\rm x}+{\beta_{\rm sh}\over \Gamma_{\rm
    sh}} \partial_y \delta u_z\right) \ , \nonumber \\ \hat{D} b_y &\,
=\, & - \partial_z \delta u_z - \frac{1}{ \Gamma_{\rm sh}^2}
\partial_x \delta u_x - {\beta_{\rm sh}\over \Gamma_{\rm sh}} \kappa
\delta u_z+ \nonumber\\ & & +u_z \biggl(-\partial_z b_y - {\beta_{\rm
    sh}\over \Gamma_{\rm sh}} \partial_z \delta u_x - {\beta_{\rm
    sh}\over \Gamma_{\rm sh}} \partial_x \delta u_z\biggr)\ ,
\nonumber \\ \hat{D} b_z &\, =\, & \partial_y \delta u_{\rm z} +
\kappa b_x + u_z \biggl(-\partial_z b_z +{\beta_{\rm sh}\over
  \Gamma_{\rm sh}} \partial_y \delta u_x\biggr)\ , \nonumber\\ \hat{D}
\delta u_x & = & -\kappa \delta u_z -\,u_z \left(\kappa b_y +
\beta_{\rm sh}{ \kappa \over \Gamma_{\rm sh}} \delta u_x +
\partial_z \delta u_x\right)\ , \nonumber \\ \hat{D} \delta u_y & = &
\beta_{\rm sh} \Gamma_{\rm sh} \kappa\, b_z + u_z\left(\kappa b_x
-\partial_z\delta u_y\right)\ , \nonumber \\ \hat{D} \delta u_z & = &
- \beta_{\rm sh}\Gamma_{\rm sh}\kappa b_{\rm y} +{\kappa\over
  \Gamma_{\rm sh}^2}\delta u_x + \beta_{\rm sh} \Gamma_{\rm sh}
\kappa\frac{\delta w}{ w} + u_z\left({\beta_{\rm sh} \over \Gamma_{\rm
    sh}} \kappa \delta u_z - \partial_z\delta u_z\right)\ , \nonumber
\\ \hat{D} \frac{\delta w}{w} &=& -\partial_x \delta u_x -\partial_y
\delta u_y -\partial_z \delta u_z \ .
\end{eqnarray}

To analyze the instabilities, it is convenient to make the following
reduction: $\hat D = k_x\beta_{\rm sh}\Gamma_{\rm sh} \tilde D$,
$\delta u_j= \beta_{\rm sh}\Gamma_{\rm sh} \tilde u_j$; $b_j$ and
${\delta w \over w}$ unchanged. The ordering for doing simplifications
is consistent with the expected physics: we assume $k_y, k_z, k_* \ll
k_x$, $u_z \ll \beta_{\rm sh}\Gamma_{\rm sh}$ and $\tilde Du_z = -
\Gamma_{\rm sh}k_*/ k_x$ at the same order than previous terms. The
simplified system reads:
\begin{eqnarray}
\tilde D b_x & \simeq & 0 \\ \tilde D b_y & \simeq &
-\frac{\partial_z}{k_x} \tilde u_z -\beta_{\rm sh} {k_* \over k_x}
\tilde u_z - \beta_{\rm sh}^2 {u_z \over \beta_{\rm sh}\Gamma_{\rm
    sh}} \frac{\partial_x}{k_x} \tilde u_z \\ \tilde D b_z & \simeq &
\frac{\partial_y}{k_x} \tilde u_z \\ \tilde D \tilde u_x & \simeq & -
     {1\over \beta_{\rm sh}} {k_* \over k_x} \tilde u_z \\ \tilde D
     \tilde u_y & \simeq & {1\over \beta_{\rm sh}} {k_* \over k_x} b_z
     \\ \tilde D \tilde u_z & \simeq & -{1\over \beta_{\rm sh}}
        {k_*\over k_x} b_y + {1\over \beta_{\rm sh}} {k_* \over k_x}
        {\delta w \over w} \\ \tilde D {\delta w \over w} & \simeq &
        -\left(\frac{\partial_x}{k_x} \tilde u_x + \frac{\partial_y}{k_x}
        \tilde u_y + \frac{\partial_z}{k_x} \tilde
        u_z\right)\label{eq:redsys}
\end{eqnarray}

Thus one gets the intermediate system:

\begin{eqnarray}
\tilde D^2 \tilde u_z &\, =\, & \left(\frac{k_*^2}{k_x^2}+\frac{k_*
  \partial_z}{\beta_{\rm sh} k_x^2} + \beta_{\rm sh} \frac{u_z}{u_{\rm
    sh}}\frac{k_* \partial_x}{k_x^2}\right)\tilde u_z +
\frac{k_*}{\beta_{\rm sh} k_x} \tilde D {\delta w \over w} \nonumber
\\ \tilde D^3 {\delta w \over w} &\, =\, & \frac{k_*\partial_x}{\beta_{\rm
    sh} k_x^2} \tilde D \tilde u_z - \frac{k_*\partial_y^2}{\beta_{\rm
    sh} k_x^3} \tilde u_z - \frac{\partial_z}{k_x} \tilde D^2 \tilde
u_z
\end{eqnarray}

\end{document}